\documentclass[aps,prl,onecolumn,showpacs,showkeys,preprintnumbers]{revtex4}
\usepackage{eurosym}
\usepackage{amsmath}
\usepackage{dcolumn}
\usepackage{bm}
\usepackage{subfigure}
\usepackage{amsfonts}
\usepackage{amssymb}
\usepackage{makeidx}
\usepackage{epsfig}
\usepackage{graphicx}

\begin{document}

\title{\textbf{The non-standad logic of physics: the case of the
Boltzmann-Sinai hard-sphere system}}
\author{Massimo Tessarotto}
\affiliation{Department of Mathematics, Informatics and Geosciences, University of
Trieste, Via Valerio 12, 34127 Trieste, Italy}
\affiliation{Research Center for Theoretical Physics and Astrophysics, Institute of
Physics, Silesian University in Opava, Bezru\v{c}ovo n\'{a}m.13, CZ-74601
Opava, Czech Republic\\
Email: maxtextss@gmail.com}
\author{Claudio Cremaschini}
\affiliation{Research Center for Theoretical Physics and Astrophysics, Institute of
Physics, Silesian University in Opava, Bezru\v{c}ovo n\'{a}m.13, CZ-74601
Opava, Czech Republic\\
Email: claudiocremaschini@gmail.com}
\author{Claudio Asci}
\affiliation{Department of Mathematics, Informatics and Geosciences, University of
Trieste, Via Valerio 12, 34127 Trieste, Italy\\
Email: casci@units.it}
\author{Alessandro Soranzo}
\affiliation{Department of Mathematics, Informatics and Geosciences, University of
Trieste, Via Valerio 12, 34127 Trieste, Italy\\
Email: soranzo@units.it}
\author{Marco Tessarotto}
\affiliation{Central Directorate for Labor, Training, Education, and Family, Permanent
Structure for Innovation and Procedure Automation, Via San Francesco 37,
Trieste, Italy\\
Email: marcotts@gmail.com}
\author{Gino Tironi}
\affiliation{Department of Mathematics, Informatics and Geosciences, University of
Trieste, Via Valerio 12, 34127 Trieste, Italy\\
Email: gino.tironi@gmail.com}
\date{\today }

\begin{abstract}
One of the most challenging and fascinating issues in mathematical and
theoretical physics concerns identifying the common logic, if any, which
underlies the physical world. More precisely, this involves the search of
the possibly-unique axiomatic logical proposition calculus to apply
simultaneously both to classical and quantum realms of physics and to be
consistent with the corresponding mathematical and filosophysical setups.
Based on the recent establishment of quantum logic, which has been shown to
apply both to Quantum Mechanics and Quantum Gravity, the crucial remaining
step involves the identification of the appropriate axiomatic logical
proposition calculus to be associated with Classical Mechanics. In this
paper the issue is posed for a fundamental example of Classical Mechanics,
which is represented by the so-called Boltzmann-Sinai dynamical system. This
is realized by the ensemble of classical smooth hard-spheres, which is set
at the basis of Classical Statistical Mechanics and is also commonly
regarded as a possible realization of Classical Newtonian Cosmology.
Depending on the initial conditions which are prescribed for such a system,
its classical state is shown to obey the propositional calculus of
non-classical logic. In particular, the latter is expressed by the 3-way
Principle of Non-Contradiction, namely the same logical principle that holds
for quantum logic. The result therefore permits to question on a
mathematical basis the principles of deterministic classical logic and the
validity of their character within the domain of Classical Physics. Such a
conclusion represents a potential notable innovation in the logical dicotomy
true/false, a crucial topic which has crossed millennia through philosophy,
logic, mathematics and physics.
\end{abstract}

\maketitle

\section{1 - Introduction}

A fundamental issue concerns the search non-standard logic of physics, with
particular reference to the boolean logic of truth propositional values,
previously treated by some of us in the context of non-relativistic quantum
mechanics (QM) and quantum gravity (GR) \cite{LOGIC-1,LOGIC-2}. The subject
of the present paper is about the search of logic for\ classical mechanics
(CM), its comparison with the quantum logic of QM and QG and the attempt to
establish a common principle of non contradiction holding both for classical
and quantum physics.

The task requires, first of all, identifying the relevant logical "truth"
propositions which are admissible for classical systems, propositions which
can be uniquely associated with the properties of their motion and state,
respectively defined by the applications "motion" and "state", namely%
\begin{equation}
\left\{
\begin{array}{c}
t\rightarrow \mathbf{r}(t), \\
t\rightarrow \mathbf{x}(t)\equiv \left\{ \mathbf{r}(t),\mathbf{v}(t)\right\}
,%
\end{array}%
\right.
\end{equation}%
with $\mathbf{r}(t)\equiv (\mathbf{r}_{1}(t),\mathbf{r}_{2}(t),..\mathbf{r}%
_{N}(t))$ and $\mathbf{v}(t)\equiv (\mathbf{v}_{1}(t),\mathbf{v}_{2}(t),..%
\mathbf{v}_{N}(t))$ being the configuration and velocity vectors, with $%
\mathbf{r}_{1}(t),\mathbf{r}_{2}(t),..\mathbf{r}_{N}(t)$ and $\mathbf{v}%
_{1}(t),\mathbf{v}_{2}(t),..\mathbf{v}_{N}(t)$ being suitably-defined
position vectors (each one here assumed to span for definiteness a suitable
subset of the same 3-dimensional Euclidean space $%
\mathbb{R}
^{3}$) and corresponding velocities.

The notable fact that we wish to establish is that, contrary to popular
belief, \ the logic of CM is not provided by classical logic (CL). The goal
is achieved by investigating a particle system of special interest in
mathematical physics, namely formed by suitable finite-sized rigid bodies
which are realized by massive rigid spheres. CM\ is usually taken to
describe\ classical particle systems represented by ensembles of massive
point particles, i.e., treated in the framework of the so-called Newton's
classical cosmology (see below), but more generally it may concern the
dynamics of arbitrary classical systems formed by ensembles of finite-size
classical rigid body systems\ obeying the CM laws due to Galilei and Newton
and are subject to mutual classical interactions such as "hard" collisions,
i.e., instantaneous elastic collisions occurring among the same rigid
bodies. An example of such systems is provided by the so-called\ \emph{%
Boltzmann-Sinai system} (B-S system; a terminology introduced in Refs. \cite%
{CMFD-T,CMFD-T-1}) which is represented by an ensemble of massive, smooth
and hard (i.e., rigid) spheres subject to unary, binary and multiple
instantaneous elastic collisions and immersed in a bounded subset of the
3-dimensional Euclidean space (see description below).

In the paper we intend to show that, provided suitable stochastic collision
laws are adopted for the B-S hard-sphere system - which is referred to for
this reason as stochastic B-S system (\emph{S-B-S system}) - its logic, and
therefore also that of CM, becomes non-deterministic, i.e., it includes also
a third truth logical value denoted as "\emph{undecidable}" (or "\emph{%
indeterminate}"). Thus, it coincides necessarily with that of a $3-$valued
propositional calculus.

In particular, in reference to the aim or the present paper our goals amount
to investigate the following issues:

\begin{itemize}
\item \textbf{First }(\emph{multiple collisions}): that, in the case of the
S-B-S system, multiple collisions may occur. In particular these include
so-called double collisions, in which two single-collisions\ (i.e.,
collisions between two hard-spheres) occur simultaneously, can generally
give rise to non-unique and therefore stochastic (i.e., multiple) particle
trajectories. As a consequence for the S-B-S system, the applicability of CL
should be necessarily ruled out, since the latter requires the validity of
the deterministic principle (i.e., uniqueness of the applications motion and
state as well as uniqueness of the configuration-space and phase-space
trajectories).

\item \textbf{Second} (\emph{non-standard logic}): that, for the S-B-S
system in place of CL some kind of non-standard logic should supersede. In
particular, besides the logical truth values \emph{true} (unique) and \emph{%
false}, a third one should occur which is referred to here as \textbf{"}%
\emph{undecidable}"(or "\emph{indeterminate}\textbf{"\ }). In other words
this means that the non-standard logic should be governed by a so-called $3-$%
\emph{way Principle of non-contradiction}\textbf{\ }(\emph{3-way PNC, }see
below in section 5) in which the occurrence of \emph{undecidability} (or
\emph{indeterminacy})\ in CM arises due to the existence of multiple, and
therefore necessarily stochastic, trajectories.\textbf{\ }

\item \textbf{Third} (\emph{analogy with quantum logic}): the undecidability
feature is remarkably similar to what happens, although for quite different
physical motivations and causes, in the case of quantum logic, i.e., the
formulation of the propositional (Boolean) logic of quantum physics. In
fact, the said logical principle should agree with the analogous $3-$way
logical principle previously determined in the two cases represented
respectively by non-relativistic QM \cite{LOGIC-1} and QG\ \cite{LOGIC-2}.

\item \textbf{Fourth} (\emph{notions of undecidability}): it should be noted
additionally that the notions of\textbf{\ }\emph{undecidability} which arise
in such contexts are potentially different from each other in the various
cases considered and, in particular, also different from that occurring in
mathematics, namely in the context of G\"{o}del undecidability theorems \cite%
{Goedel}. In fact, in mathematics \ undecidability usually occurs when and
an indeterminate form\ arises, namely a mathematical expression occurs whose
value cannot be determined (i.e., for example, if it does not admit an
unique value as in the case of trajectories for CM) and, respectively in QM,
QG and \ for the B-S system. In fact, while as shown in QM and QG \cite%
{LOGIC-1,LOGIC-2} this is due to the possible occurrence of observables
which are endowed with an infinite standard deviation, instead, in the
context of the B-S system undecidability occurs - as we intend to prove here
- because of the occurrence of stochastic trajectories.

\item \textbf{Fifth} (\emph{connection with G\"{o}del undecidability theorems%
}): finally, the connection with G\"{o}del undecidability theorems is
investigated and two issues are addressed: a) whether the notion of
undecidability available in the context of the B-S system is consistent or
not with that adopted in the context of G\"{o}del theory; and b) whether the
B-S system may represent or not an example-case and an application of G\"{o}%
del undecidability theorems \cite{Goedel}.
\end{itemize}

\section{2 - The state of the art}

This work is part of an investigation dealing\textbf{,} on one hand, with
the search of logic in physics and, on the other hand, with the mathematical
properties of the B-S hard-sphere system.

Regarding logic, here we consider in particular the case of CM. Customarily,
this refers to the well known formulation of Newton's classical cosmology
based on the deterministic point particle model of universe, to be
considered as a finite/infinite but numerable ensemble of massive point
particles moving in a flat Galileian spacetime.\ In fact, its connection
with classical logic is well-known. This was first recognized by Immanuel
Kant in his\ \emph{Critique of Pure Reason} \cite{Kant,ONLINE-IV-A}, who
remarked that based on Newton's principle of determinism \cite{Newton},
Newton's classical cosmology could only obey the 2-way logic of true and
false propositions, i.e., classical logic, and therefore satisfy the
Aristotle's 2-way PNC.

However, the deterministic description of classical matter afforded by
Newtonian cosmology may not hold for more general body systems, such as in
particular the B-S system. Nevertheless, the same subject matter (i.e.,
logic) pertains more generally both to classical and quantum physics. The
subject has been recently investigated in the context of non-relativistic
quantum mechanics and quantum gravity, for which a common $3-$way Principle
of non-contradiction has been proved to hold (see again \cite%
{LOGIC-1,LOGIC-2}). Finally, as far as the B-S hard-sphere system is
concerned, its mathematical properties must carefully be dealt with also in
view of the considerable amount of related specialized literature available
in the field and dealing specifically with statistical mechanics and its
kinetic theory treatment.

\subsection{2.1 - The classical logic of Newton's classical cosmology}

It is usually regarded as part of the common wisdom that CM should be
deterministic (i.e., it should obey the so-called Newton's deterministic
principle, see below) and hence should be governed by CL.\textbf{\ }

According to this view, the rationale should therefore be founded on the $2-$%
valued propositional calculus (i.e., in which the only logical truth values
are\emph{\ true} and \emph{false}), fulfilling the principles of CL, namely,
the Law of identity or deterministic principle (any true trajectory is
unique), the Law of non-contradiction (any true proposition (trajectory)
cannot be false) and the Law of excluded middle (any proposition
(trajectory) is either true or false, no other possibility arises). These
three laws can be summarized respectively by

\begin{itemize}
\item The 2-way principle of non-contradiction (2-way PNC), namely in which
the only possible logical and mutually exclusive\ truth values are true or
false;

\item And the deterministic principle (any true logical proposition is
unique).
\end{itemize}

Therefore, in CM the true logical value (which identify the semantics of CM)
should be uniquely associated with the deterministic state of a classical
system. This must be identified with a vector field\ of the form $\mathbf{x}%
(t)\equiv (\mathbf{r}(t),\mathbf{v}(t))$,\ in which $\mathbf{r}(t)$\ and $%
\mathbf{v}(t)$\ denote the position and velocity vectors prescribing its
state.

The traditional approach for the explicit construction of the 2-way semantic
true/false is based on the\ same strategy originally pointed out by Birkhoff
and von Neumann \cite{Birkhoff-vonNeumann} (B-vN) for Hamiltonian mechanics.
Accordingly, in CM the truth values, which are attached to all possible
phase-space trajectories (and which by assumption\ should realize at least $%
C^{(0)}$, i.e., continuous, configuration-space trajectories), are uniquely
identified as follows:

\begin{itemize}
\item 1st logical truth proposition\ "\emph{true}": in CM, for a prescribed
initial state\ $\mathbf{x}(t_{o})\equiv \mathbf{x}_{o}\equiv (\mathbf{r}%
(t_{o})=\mathbf{r}_{o},\mathbf{v}(t_{o})=\mathbf{v}_{o})$, we shall denote
as "true" a phase-space trajectory, if it coincides with that determined by
Newton's equations, namely the trajectory generated by the corresponding
classical flow defined for all $t,t_{o}\in I$\ ($\subseteq
\mathbb{R}
$),%
\begin{equation}
\mathbf{x}(t_{o})\equiv \mathbf{x}_{o}\rightarrow \mathbf{x}(t)\equiv \chi (%
\mathbf{x}_{o},t_{o},t),  \label{flow}
\end{equation}%
which relates the initial state $\mathbf{x}(t_{o})\equiv \mathbf{x}_{o}$\
with its image at time $t$. This is determined by solving the Cauchy
initial-value problem associated with the corresponding Newton equations of
motion for the same classical system.

\item 2nd logical truth proposition "\emph{false}": again, given the
prescribed initial state $\mathbf{x}(t_{o})\equiv \mathbf{x}_{o}\equiv (%
\mathbf{r}(t_{o})=\mathbf{r}_{o},\mathbf{v}(t_{o})=\mathbf{v}_{o})$, we
shall denote as "false" any other trajectory corresponding to different
initial conditions, namely of the type%
\begin{equation}
\mathbf{x}(t_{o})\equiv \mathbf{x}_{o}\rightarrow \mathbf{x}^{\prime
}(t)\equiv \chi (\mathbf{x}_{o}^{\prime },t_{o},t),
\end{equation}%
where $\mathbf{x}_{o}^{\prime }\neq \mathbf{x}_{o}$.
\end{itemize}

In this regard, a\ well known example is provided by Newtonian (Classical)
cosmology (see Ref. \cite{Hetherington} and also Appendix A in Ref. \cite%
{LOGIC-1}). This is based on the representation of matter in terms of a
finite or numerable set of massive point particles which are immersed in a
so-called Galileian space-time (i.e., the universe). The latter is by
assumption taken to be the direct product of an infinite Euclidean $3-$%
dimensional configuration space (identified with the ambient space) and an
infinite Euclidean $1-$dimensional time axis (in which therefore time is
"absolute", i.e., frame-independent). The point particles obey the Newton
equations of motion, an assumption which is taken to hold identically, i.e.,
at all times and in the whole space, except at discrete "collision" events
and discrete positions of configurations space where mutual (unary, binary
or multiple) instantaneous collisions occur. In the case of binary and
multiple collisions, the latter are also taken to be elastic, i.e., to
conserve total linear momentum and total kinetic energy across collisions.
Hence, it follows that both binary and multiple collision laws are
necessarily deterministic. In the case of point particles this is an obvious
consequence of the assumed collision law. In fact, considering $k\geq 2$\
colliding point particles which are referred, without loss of generality, to
the ensemble center of mass (CM) for the colliding particles. It follows
that, if the initial (incoming before collision) and final (outgoing after
collision) states are denoted respectively by\ $\left\{ (\mathbf{r},\mathbf{v%
}^{(-)i},i=1,k\right\} $\ and $\left\{ (\mathbf{r},\mathbf{v}%
^{(+)i},i=1,k\right\} $\ (relative to the said CM of colliding particles),
the so-called\textbf{\ }\emph{bounce-backward reflection collision law}%
\textbf{\ }applies, according to which:%
\begin{equation}
\mathbf{v}^{(+)i}=-\mathbf{v}^{(-)i},  \label{eq-1}
\end{equation}%
i.e., after collision the (outgoing) velocities of all colliding particles $%
\left\{ (\mathbf{r},\mathbf{v}^{(+)i},i=1,N\right\} $ are "bounced backward"
with opposite directions with respect to the incoming velocities of the same
particles, both defined in the CM of all the particles taking part in the
collision event. It is obvious that the collision law (\ref{eq-1}) conserves
the relevant collisional invariants defined in the same CM, namely the
CM-total kinetic energy and the CM-total linear momentum of the sane
colliding particles.

This confirms therefore that Newtonian cosmology is deterministic. Namely,
that provided the universe were indeed fulfilling Newtonian cosmology,%
\textbf{\ }"\emph{We ought then to regard the present state of the universe
as the effect of its anterior state and as the cause of the one which is to
follow}." This is indeed the famous sentence expressed by Pierre Simon de
Laplace in one of his latest works \cite{Laplace}. Which represents of
course the view first expressed by Isaac Newton himself (1658 \cite{Newton}%
), stressed in some sense stronger sense by Ruggero Boskovich, the Dalmatian
astronomer and scientist from Ragusa (1758 \cite{Boskovich}) and later
invoked by Immanuel Kant in his theorization of the philosophy of Illuminism
(Enlightenment, 1781 \cite{Barrow}).

However, the same deterministic collision law may not hold for more general
(i.e., in particular finite-size) particle systems and, consequently, the
conventional $2-$way classical logic may not hold for them. The task of the
present investigation is to go beyond the limitations set by Newtonian
cosmology. That is, to prove that - contrary to what implied by Newtonian
cosmology - the logic of CM is not two-valued.

The result is achieved by investigating the $N-$body Boltzmann-Sinai
hard-sphere system, namely the $N-$body billiard problem formed by smooth
hard spheres immersed in a rigid parallelepiped-shaped and having stationary
boundary. The same particles, besides unary instantaneous elastic collisions
with the boundary, are assumed to undergo instantaneous elastic binary and
multiple collisions among them.

Based on the fact that multiple collisions for this system may not generally
give rise to a unique outgoing state\ (i.e., defined after multiple
collisions), and therefore be deterministic in character, we intend to show
that multiple collision laws are generally stochastic, in the sense that
they may correspond to different - but equally probable - sequences of
simultaneous binary\ collisions events. This amounts to determine a
"statistical prescription" on the Boltzmann-Sinai hard-sphere system and to
analyze its basic qualitative properties, with special reference to the
determination of its logic.

\subsection{2.2 - A more general problem}

As already anticipated, however, the issue concerns, in principle, the
broader perspective of identifying the logic of physics in general, i.e.,
both classical and quantum physics, where the latter should include quantum
gravity besides quantum mechanics.

The logic of QM and the logic of QG have actually been regarded for a long
time fundamental as unsolved problems of mathematical and theoretical
physics. The starting inspiration was provided in 1936 by a seminal paper,
titled The Logic of Quantum Mechanics, due to two legendary scientists,
famous for their outstanding contributions in several areas of Quantum and
Statistical Mechanics, namely Garrett Birkhoff and John von Neumann \cite%
{Birkhoff-vonNeumann}. This paper is usually accredited for the formulation
of the so-called \emph{standard }(approach to)\emph{\ quantum logic.}
However, despite this credited title (and the side contribution, cited
above, for establishing also the rules for connecting Hamiltonian Mechanics
with CL) it did not actually solve the problem,\ at least for the following
main reasons:

\begin{enumerate}
\item First, standard quantum logic is not able to recover a boolean logic,
and - in particular -\ does not identify the appropriate notions of true and
false propositions to be based on the notions quantum expectation values and
standard deviations available in standard QM.

\item Second, the possible identification of the undecidability (or
indeterminacy) logical truth value, relevant for establishing a possible
realization of the G\"{o}del incompleteness theorems \cite{Goedel}, was
actually never achieved.

\item Third, the same standard quantum logic does not explicitly take into
account the Heisenberg inequalities, together with the generalized ones
later discovered in Ref. \cite{CMFD-T-22}.

\item Fourth, it did not include an analogous treatment for quantum gravity
(QG) where analogous generalized Heisenberg inequalities can be shown to
hold (see the above reference).

\item Later on, however, further issues would have emerged, some of them
specifically related to QG. In fact in the context of QG the generalized
Heisenberg inequalities should take an explicit $4-$tensor form, implying in
turn that the setting of quantum gravity should be established in the
context of a $4-$tensor covariant representation (see Refs. \cite%
{cqg-3,cqg-4}).\
\end{enumerate}

The first two are most likely the reasons why, at the time, von Neumann was
apparently unsatisfied by the outcome of their paper (quantum gravity was
still long time into the future). In fact, despite the fact that von Neumann
was well aware of the undecidability theorems \cite{Goedel} established by\ G%
\"{o}del already in 1931 (he personally attended the same International
Conference of Mathematics in K\"{o}nigsburg where G\"{o}del had first
presented his famous theorems), in his work with Birkhoff he missed any
possible link with the logic of propositional calculus based on Boolean
logic. And consequently also with G\"{o}del's theorems.

Nevertheless, in a sense, the same paper was also very successful. In fact,
it was responsible for the beginning of a formidable and long-lasting
conceptual debate on the subject of quantum logic (QL). Notably, this
happened especially in the broader community of mathematicians and
mathematical physicists \cite{Jammer,Corsia}, becoming actually a full part
of Mathematical Logic, Model Theory and Topology \cite{Chang-Keisler}.
Indeed, among mathematicians the subject developed in a full-fledged new
research topic, largely independent of the original ideas and problems
raised by B-vN. It has to be noted in this regard that this goes into the
direction of the attempt to formulate a new logic for QM. An attempt which
has been under way until very recent times (see in particular \cite%
{Ellermann2024}), in accordance with the view that propositional logic
(i.e., the logic of admissible truth values) should be somehow replaced by
the "logic of partitions (or equivalence relations)" \cite{Ellerrmann2010}.\
Such an approach, however, just as the original contribution due to B-vN,
seems incomplete because, for example, it is unrelated to classical logic
even in the so-called semiclassical limit. Nevertheless the quantum logic
they (B-vN) devise is apparently multi-valued.\textbf{\ }From this point of
view there it appears not completely new. In this regards, in fact, it has
to be noted that already from the nineteen century multi-value logic has
been considered by mathematicians \cite{Boole} (for a historical review see
also \cite{Church}), its foundation being based on multi-valued
propositional calculus \cite{Saunders}.

However, the logical truth values true/false in case of the B-vN paper
cannot be defined at all, even if the logic of classical Hamiltonian systems
can apparently be correctly recovered in the semiclassical limit (indeed, in
the context of the semiclassical approximation for QM, projection operators
are analogous to the characteristic functions defined on subsets of
appropriate phase spaces \cite{Lafleche}).$\ $Indeed, B-vN thought that the
appropriate propositional calculus for quantum mechanics might be achieved
by identifying the logical propositions with the projections of the physical
observables.

The same idea actually had emerged already in earlier paper. In fact, in his
classic treatise \cite{Von Neumann} John von Neumann had already pointed out
that projections on a Hilbert space might be viewed as propositions about
physical observables. Nevertheless, despite a few attempts \cite{Mackay}
(see also \cite{Ya'acov} where a comparison with G\"{o}del undecidability
theorems was suggested but never actually achieved), such a choice - which
von Neumann himself later seemed to consider unsatisfactory - was the reason
why several further attempts to the formulation of quantum logic were
proposed (see for example \cite{Doering}). To the point that nowadays in
some texts the term \textquotedblleft quantum logic\textquotedblright\ is
used rather narrowly to refer to the algebraic and order-theoretic aspect of
propositional logic. More detailed introduction can be found in the books of
Beltrametti and Cassinelli \cite{Beltrametti}, Kalmbach \cite{Kalmbach},
Cohen \cite{Cohen}, Pt\'{a}k and Pulmannov\'{a} \cite{Ptak}, Giuntini \cite%
{Giuntini} and Svozil \cite{Svozil,Svozil-1} among others.

Such a type of approaches, interesting as they may be, still have remained
incomplete. One possible reason is that the formulation of the logic of QM
and QG requires, foremost, a careful investigation on the conceptual
foundations of physics, i.e., specifically QM, QG and even CM,\ including
the possible analogies between classical and quantum physics \cite%
{Eyal-2,Eyal-1}\textbf{.}\ In fact, among the questions that arise when
considering the problem of logic\ in physics, one is represented by the same
reason for which these issues have remained unsolved for such a long time in
different fields of physics. This concerns, more precisely, the possible
identification of a \emph{unique and common logical principle}\textbf{, }%
i.e., based on the same kind of Principle of non-contradiction,\ \emph{%
together with the same discrete logical truth values}\textbf{, }among
different branches of physics, which should include, besides QM and QG, also
CM.

However, for the first two disciplines a possible candidate for the
identification of quantum logic in terms of a $3-$valued propositional
calculus, which obeys a corresponding $3-$valued Principle of
Non-Contradiction, has actually recently been proposed \cite{LOGIC-1,LOGIC-2}%
. The result, striking for its simplicity, is related to the conditions of
validity of the Heisenberg uncertainty principle and also a proper
definition of the notion of admissible\ logical\ truth values arising in the
context of quantum theories.\ More precisely, such a principle, which is
intrinsic to all quantum theories, establishes a relationship between the
quantum standard deviations holding between conjugate quantum variables (to
be regarded in all cases as quantum observables). Nevertheless such a
principle requires for its validity that the standard deviations are
non-zero nor infinite. However when they are infinite the corresponding
observable is obviously indeterminate (in the sense that its quantum
expectation value is characterized by an infinite quantum error
measurement)! This means that observables can be in some cases
indeterminate. As a consequence, depending on the dynamical evolution of the
quantum state and on the physically admissible standard deviations (namely
those which are compatible with the phenomenon of the quantum minimum length
\cite{Tex2023}), three possible truth logical values can be established,
which are labeled to as \textit{true,} \textit{false} and \textit{%
indeterminate }(here equivalently referred to as \textit{undecidable)}. In
particular, for a given quantum observable, \textit{truth} (a), \textit{%
falsity} (b) or \textit{undecidability} (or equivalently \textit{%
indeterminacy)} (c), respectively occur:

a)-b) if its time\ evolution is consistent (true) or not (false) with the
time evolution of the quantum state and if its quantum standard deviations
remain finite (even zero) at all times;

c) if the standard deviation of a quantum observable becomes infinite.

However, QL should exhibit a further key property. More precisely, QL should
recover exactly CL in the (appropriate) semiclassical limit in which,
respectively:

\begin{itemize}
\item QM recovers CM;

\item QG recovers classical General Relativity.
\end{itemize}

To enable such features, a suitable representation of QM and, respectively,
of QG, has been introduced in the cited references. This is based on the
adoption of appropriate \emph{stochastic-trajectory representations,} i.e.,
introducing stochastic Lagrangian trajectories - denoted as Generalized
Lagrangian Paths (GLP) - which, in each case, span a corresponding
configurations space. There are three main reasons for choosing such a type
of representation:

\begin{itemize}
\item The first reason is that Lagrangian trajectories (both in QM and QG)
should be stochastic in nature for consistency with the requirement that
quantum particle trajectories are by definition stochastic.

\item The second reason is that, as shown in Refs. \cite{GLP-2016} and \cite%
{cqg-6}, this feature allows to construct explicit analytical solutions of
the corresponding quantum-wave equation.

\item In turn, the same feature becomes of crucial importance also for the
explicit analytic evaluation of the relevant quantum standard deviations
which are needed for the establishment of quantum logic.
\end{itemize}

The purpose of the present investigation - representing, in a sense, the
completion of the previous one given in \cite{LOGIC-1,LOGIC-2} - is to
address the same fundamental issue also in the context of classical
mechanics, thus posing the problem in a broader perspective, i.e., relevant
for the establishment of the \emph{logic of physics in general}. We
anticipate that, rather remarkably, an analogous feature, based on the
adoption of \emph{stochatic-trajectories}, is found to characterize also the
construction of the logic of CM.

In this regard, famous is Immanuel Kant's \emph{declaration of faith} \cite%
{Kant} in favour of Newtonian cosmology and hence, implicitly, of CM itself.
As stated above, this is closely related to the Newton principle of
determinism, i.e., the uniqueness of classical trajectories, characteristic
of CM. Such a statement is expressed in his book (1781) "\emph{Critique of
Pure Reason}" \cite{Kant} where he stated that the validity of Newtonian
Cosmology \cite{Newton} should never be questioned, since, in his own words,
"\textit{the philosopher is not [should not be, EN] permitted to avail [a]
possible experience [that] is not what can be given in an experiment 'in
concreto' but what is contained in the idea to which the empirical synthesis
must approximate}". Which clearly means that Kant regarded Newtonian
cosmology as an absolute truth, consistent with the blind faith in science,
characteristic of Illuminism \cite{Barrow}.

Paradoxically, however, the actual association of CM with classical logic
must first be ascribed to Birkhoff and von Neumann. In fact, their paper
\cite{Birkhoff-vonNeumann} is the first one to point out the rigorous
connection existing between Hamiltonian dynamics and classical logic (CL).
However, Hamiltonian systems represent only a particular subset of CM which
includes in principle also many-body and finite-size systems, such as
hard-sphere systems which may be subject to hard instantaneous collisions. A
particular case is provided by the $N-$body Boltzmann-Sinai hard-sphere
system\ which is constituted by an ensemble of $N>2$ smooth hard spheres
subject to instantaneous elastic hard collisions and in which, by
definition, the same particles do not exchange angular momentum, either
during arbitrary collisions with other particles or with the boundary. We
stress that none of the implications considered in this research arise in an
"universe" of $N\leq 2$\ particles as hard spheres. Remarkably, the ancient
philosophers, like Laozi \cite{Laozi}, had the intuition that the number 3
generates the variety of universe (while number $1$\ simply produces number $%
2$,\ for example by negation).

\subsection{2.3 - Previous investigations on the Boltzmann-Sinai hard-sphere
system}

The choice of the Boltzmann-Sinai system, which is intrinsically
non-Hamiltonian because hard collisions are ordinarily not susceptible of an
Hamiltonian treatment, seems also well motivated from another viewpoint,
i.e., because it is one of the abstract physical systems which has more
extensively been investigated in the last 150 years. In fact, the
Boltzmann-Sinai system is the same one for which:

\begin{itemize}
\item The property of ergodicity on the energy-hypersurface phase-space was
first rigorously established by Sinai (1963-1970) for the case $N=2$\ \cite%
{Sinai1963,Sinai1970,Sinai1987}, then for $N=3$\ and $N=4$\ hard spheres
\cite{KSS91,KSS92}. However, for $N>4$\ technical difficulties arise for the
step for $N$\ to $N+1$\ hard spheres \cite{BLPS92} (see the related review
given in Ref. \cite{BGST20}).

\item The asymptotic kinetic theory of rarefied gases was first established
in 1872 by Ludwig Boltzmann \cite{Boltzmann1872} based on his namesake
kinetic equation and by Harold Grad \cite{Grad}, who established its
connection with classical statistical mechanics. The Boltzmann approach,
based on the treatment of smooth-hard-spheres binary collisions, has been
proved correct for smooth PDFs \cite{CMFD-T-3}, despite the adoption (for
the $2-$body PDF)\ of physically-incorrect collision boundary conditions
(CBC) in terms of the so-called PDF-conserving CBC \cite{CMFD-T-4}.

\item The limits of validity of the Boltzmann equation and related H-theorem
were first established \cite{CMFD-T} showing that both become incorrect in
the case of the so-called "certainty" PDF, i.e., the Dirac delta PDF.

\item The exact statistical theory for finite systems of smooth hard
spheres, subject to binary and unary collisions, was established for the
first time in terms of the so-called Master kinetic equation \cite{CMFD-T-1}%
. This was achieved thanks to the discovery of a physically-based modified
CBC for the $N-$body PDF \cite{CMFD-T-4, CMFD-T-4BIS}, which differs
intrinsically from the PDF-conserving CBC adopted originally by Boltzmann.
\end{itemize}

In particular, the same Master kinetic equation was shown:

\begin{itemize}
\item To imply, in contrast to the Boltzmann H-theorem, the validity of a
constant-H theorem\ \cite{CMFD-T-2BIS}.

\item To be globally valid \cite{CMFD-T-2}.

\item Finally, to recover exactly the Boltzmann equation and the Boltzmann
H-theorem in the so-called Boltzmann-Grad limit \cite{CMFD-T-3}.
\end{itemize}

\section{3 - The open problem}

In this section the main open problems for the B-S system are addressed.
These concern the treatment of multiple collisions.

\subsection{3.1 - Treatment of multiple collisions in smooth-hard-sphere
systems}

The aim of the subsequent part of the paper is to investigate two issues:

\begin{itemize}
\item The prescription of multiple collisions occurring in many-body
classical systems. In the following, as an illustration, we shall limit our
analysis to double collisions only.

\item The determination of the logic of classical systems in which such
multiple elastic collisions occur.
\end{itemize}

The case considered here refers to a set of like, smooth hard spheres, i.e.,
massive rigid spheres of diameter $\sigma >0$ which do not exchange angular
momentum among them (nor with the boundary), but carry (only) linear
momentum and kinetic energy, which are exchanged among colliding particles
but in which total linear momenta and kinetic energy of the colliding
particles are conserved. This is the case of the so-called B-S system. The
latter is the ensemble of $N>2$\ like classical particles represented by
smooth hard spheres which belong to the bounded\ configuration interior
domain of a $3-$dimensional bounded subset of the Euclidean space $%
\mathbb{R}
^{3}$ (the so-called $3-$dimensional billiard),\ which is assumed to be
endowed with a stationary, rigid and infinitely massive
parallelepiped-shaped boundary. Moreover, the spacetime itself is assumed to
be embedded in a Galileian spacetime. The goal is to analyze a conceptual
aspect which is usually simply ignored in most of previous literature but
becomes of crucial importance in the present context. This is about the role
of multiple collisions for the same B-S system.

The second issue to be addressed here, namely the logic of the B-S system,
requires investigating certain qualitative properties of the B-S system\emph{%
\ per se}. Not surprisingly, this means, first of all, focusing our
attention on a crucial topic, previously completely ignored in the
literature. This concerns the treatment of multiple-collisions in
smooth-hard-spheres systems, i.e., in which three or more of such spheres
collide simultaneously. This means that they should come simultaneously into
contact, with negative relative normal speeds among colliding particles,
i.e., more precisely such that for all the colliding particles $i$ and $j$
and\ denoting by $\mathbf{v}_{i},\mathbf{v}_{j},$\textbf{\ }$\mathbf{r}_{i},%
\mathbf{r}_{j}$ and $\mathbf{n}_{ij}=(\mathbf{r}_{i}-\mathbf{r}%
_{j})/\left\vert \mathbf{r}_{i}-\mathbf{r}_{j}\right\vert $ their
velocities, positions and relative-position unit vectors for which $\left.
\mathbf{r}_{i}=\mathbf{r}_{j}+\sigma \mathbf{n}_{ij}\right. $, it occurs
that:%
\begin{equation}
\mathbf{n}_{ij}\cdot \left( \mathbf{v}_{i}-\mathbf{v}_{j}\right) <0.
\end{equation}

\subsection{3.2 - Axiomatic propositions}

In particular, we intend to prove three propositions (PROP. \#1-\#3 below)
showing that multiple collisions can be considered as realized by different
ordered sequences of instantaneous binary collisions. Such sequences
generally lead to different outgoing (i.e., after collision) states of the
colliding particles. As a consequence, multiple collision laws are generally
non-unique.

For this purpose, we first intend to show that, based on the prescription of
the B-S system, the following proposition holds:

\bigskip

\textbf{PROPOSITION \#1: multiple-collision laws are necessarily realized by
suitable ordered sequences of instantaneous elastic binary collisions.}

\emph{In the context of CM, for the B-S system the collision laws of
multiple instantaneous elastic collisions can only be realized by means of
suitable ordered sequences of instantaneous binary collisions. Thus, for
example, in the case of double instantaneous elastic binary collisions they
are actually realized by ordered sequences of three different instantaneous
elastic binary collisions.}

\emph{PROOF OF PROP. \#1}

Let us consider as a reference example the case of the so-called double
collision event $(1-2)+(2-3)$\ in which three smooth hard spheres, labeled
with "1", "2" and "3", happen to experience two simultaneous collisions
among them, namely $(1-2)$\ and $(2-3)$. The double collision occurs when
simultaneously the spheres "1" and "2" and separately the spheres "2" and
"3" collide mutually elastically. Such a condition is realized if particle
"1" and "2" and separately the spheres "2" and "3" are in an "incoming"
state (i.e., a state before a double collision event occurs), namely are
approaching each other. This requires that relative velocities of particles
"1" and "2" and separately the spheres "2" and "3", namely $\mathbf{v}_{12}=%
\mathbf{v}_{1}-\mathbf{v}_{2}$\ and $\mathbf{v}_{23}$\ $=\mathbf{v}_{2}-%
\mathbf{v}_{3},$\ satisfy the two inequalities%
\begin{eqnarray}
\mathbf{n}_{12}\cdot \mathbf{v}_{12} &<&0,  \label{12} \\
\mathbf{n}_{23}\cdot \mathbf{v}_{23} &<&0,  \label{23}
\end{eqnarray}%
with $n_{ij}$\ denoting the unit vectors $\mathbf{n}_{ij}=(\mathbf{r}_{i}-%
\mathbf{r}_{j})/\left\vert \mathbf{r}_{i}-\mathbf{r}_{j}\right\vert .$
Instead, the "outgoing" state (i.e., a state which occurs after a double
collision event) must correspond to the condition in which,
\begin{eqnarray}
\mathbf{n}_{12}\cdot \mathbf{v}_{12} &>&0,  \label{12-b} \\
\mathbf{n}_{23}\cdot \mathbf{v}_{23} &>&0.  \label{23-b}
\end{eqnarray}

We notice that, despite the two binary collisions $(1-2)$\ and $(2-3)$\
occur simultaneously, one has still to decide which of the two happens
first. In the case of double collisions there are therefore two possible
ordered sequences of simultaneous binary collisions. We anticipate that the
distinction of the various sequences makes sense because the same sequences
are physically different in the sense that generally they lead to different
outgoing states (see Propositions \#2 and \#3).

Correspondingly, two questions arise: 1) what are the required conservations
laws in multiple collisions and 2) what are the actual instantaneous elastic
binary collisions taking place between the same particles. Without loss of
generality we shall also assume that, before the collision event occurs, the
velocity of the (center of the) sphere "2" vanishes, i.e., $\mathbf{v}_{2}=0$%
\ and that it coincides also with the velocity of the center of mass (CM) of
the system of hard spheres $(1,2,3)$\ (notice nevertheless that for greater
generality we shall assume that the center of the sphere "2" may not
coincide with the same CM). Then, in this setting if $v_{1}$\ is the initial
velocity of the center of sphere "1", the corresponding velocity of the
center of sphere "3" is necessarily such that $\mathbf{v}_{3}=-\mathbf{v}%
_{1} $\textbf{.}

In fact, one notices that if naively one tries to impose the conservation
laws on the two separate collisions $(1-2)$\ and $(2-3)$,\ namely

\begin{itemize}
\item The conservation of the sum of the kinetic energies and, separately,
of the linear momenta of particles "1" and "2" which should remain the same
before and after collision $(1-2)$;

\item The conservation of the sum of the kinetic energies and, separately,
of the linear momenta of particles "2" and "3" which should remain the same
before and after collision $(2-3)$;
\end{itemize}

then one can prove that the above conservation laws cannot generally be
fulfilled. In fact, to overcome the difficulty requires determining suitable
sequences of instantaneous elastic binary collisions.\ Indeed, in the case
of the double collision event $(1-2)+(2-3)$, one can immediately realize
that actually there necessarily occur not two but three distinct,
simultaneous, binary collisions. The third one in fact is a consequence of
the fact that the second particle after colliding once with the other
particles must necessarily "bounce back" and experience another collision
with the particle with which it has collided first.

Now, any double collision can be realized by means of two possible binary
collision sequences (occurring simultaneously but in a certain order). Thus,
denoting $(i-j)$\ a binary collision between particles $i$\ and $j$,\ these
sequences are realized respectively by
\begin{equation}
(1-2)+(2-3)+(1-2),  \label{case A}
\end{equation}%
or by the alternate sequence
\begin{equation}
(2-3)+(1-2)+(2-3).  \label{case B}
\end{equation}

Next, we intend to prove the following related propositions.

\begin{itemize}
\item \textbf{PROPOSITION \#2: general non-uniqueness of multiple-collision
laws}. In general, in a multiple collision occurring among $N\geq 3$ smooth
hard spheres the final velocities of the particles are not uniquely
defined.\ In the particular case of double collisions, similarly the final
velocities of the particles are generally not uniquely defined. More
precisely a double collision can be achieved by means of two non-equivalent
sequences of binary collisions. The non-uniqueness occurs when the relative
position unit vectors $\mathbf{n}_{12}$\ and $\mathbf{n}_{23}$\ (where $%
\mathbf{n}_{ij}$ $(=-\mathbf{n}_{ji})$\ denote the unit vectors joining the
centers of particles "$i$"\ and "$j$") are arbitrarily directed (namely they
are not parallel or orthogonal to each other).

\item \textbf{PROPOSITIONS \#3A and \#3B: exceptions}.\ The exception, which
corresponds to the case in which the double collision is deterministic,
i.e., the final velocities after the double collision are uniquely
determined, is represented by two possible cases (labeled respectively%
\textbf{\ }\emph{3A} and \emph{3B}). The first case (\emph{Proposition \#3A}%
) occurs when the two unit vectors\textbf{\ }$\mathbf{n}_{12}$\textbf{\ }and%
\textbf{\ }$\mathbf{n}_{23}$\textbf{\ }are parallel to each other, namely:%
\textbf{\ }%
\begin{equation}
\mathbf{n}_{12}\cdot \mathbf{n}_{23}=1,  \label{symmetry}
\end{equation}%
which corresponds to so-called symmetric initial conditions. The second case
(\emph{Proposition \#3B}) is the one in which the same unit vectors are
mutually orthogonal, namely,%
\begin{equation}
\mathbf{n}_{12}\cdot \mathbf{n}_{23}=0.  \label{orthogonal}
\end{equation}%
In this case in fact, in a proper sense, the double collision reduces to two
independent binary collisions, namely $(1-2)$\ and $(2-3)$, occurring
simultaneously.
\end{itemize}

\subsection{3.3 - Direct proof of Propositions \textbf{\#3}}

First, we provide a direct proof of Proposition \#3A. For definiteness, let
us assume that the condition of symmetry (\ref{symmetry}) is satisfied
before a double collision takes place. Furthermore, let us assume that in
the center of mass (CM) the following initial velocities are satisfied:%
\begin{equation}
\left\{
\begin{array}{c}
\mathbf{v}_{1} \\
\mathbf{v}_{2}=0 \\
\mathbf{v}_{3}=-\mathbf{v}_{1}%
\end{array}%
\right. .  \label{initial conditions}
\end{equation}%
Then, elementary symmetry considerations imply that:

\begin{itemize}
\item The hard sphere "2" remains at rest with respect to the CM, so that,
after the symmetric double collision, its outgoing velocity vanishes: $%
\mathbf{v}_{2}^{\prime \prime \prime }=0$.

\item The hard spheres "1" and "3" undergo an analogous collision law,
namely a bounce-back in the direction of symmetry which is determined by the
unit vector $\mathbf{n}_{12}=\mathbf{n}_{23}$.

\item As a consequence, the following collision law is necessarily fulfilled
during the symmetric double-collision:%
\begin{equation}
\left\{
\begin{array}{c}
\mathbf{v}_{1} \\
\mathbf{v}_{2}=0 \\
\mathbf{v}_{3}=-\mathbf{v}_{1}%
\end{array}%
\right. \rightarrow \left\{
\begin{array}{c}
\mathbf{v}_{1}^{\prime \prime \prime }=\mathbf{v}_{1}-2\mathbf{n}_{12}%
\mathbf{n}_{12}\cdot \mathbf{v}_{1} \\
\mathbf{v}_{2}^{\prime \prime \prime }=0 \\
\mathbf{v}_{3}^{\prime \prime \prime }=-\mathbf{v}_{1}+2\mathbf{n}_{12}%
\mathbf{n}_{12}\cdot \mathbf{v}_{1}%
\end{array}%
\right. .  \label{symmetric double-collision law}
\end{equation}
\end{itemize}

Next let us consider the proof of Proposition \#3B. In case of initial
velocities (\ref{initial conditions}) and in validity of the orthogonality
condition (\ref{orthogonal}), then the following collision law is
necessarily fulfilled during the symmetric double-collision:%
\begin{equation}
\left\{
\begin{array}{c}
\mathbf{v}_{1} \\
\mathbf{v}_{2}=0 \\
\mathbf{v}_{3}=-\mathbf{v}_{1}%
\end{array}%
\right. \rightarrow \left\{
\begin{array}{c}
\mathbf{v}_{1}^{\prime \prime \prime }=\mathbf{v}_{1}-\mathbf{n}_{12}\mathbf{%
n}_{12}\cdot \mathbf{v}_{1} \\
\mathbf{v}_{2}^{\prime \prime \prime }=\mathbf{n}_{21}\mathbf{n}_{21}\cdot
\mathbf{v}_{1}+\mathbf{n}_{23}\mathbf{n}_{23}\cdot \mathbf{v}_{1} \\
\mathbf{v}_{3}^{\prime \prime \prime }=-\mathbf{v}_{1}-\mathbf{n}_{23}%
\mathbf{n}_{23}\cdot \mathbf{v}_{1}%
\end{array}%
\right. .
\end{equation}

\subsection{3.4 - Proof of Proposition 2}

The proof of Proposition \#2 is now obtained by constructing explicitly the
possible sequences of binary collisions. As stated above, the initial
velocities are in both cases taken to fulfill (\ref{initial conditions}).

\subsubsection{First case: sequence (1-2)+(2-3)+(1-2)}

Let us consider for this purpose the first sequence (\ref{case A}). Then,
denoting by $\mathbf{n}_{ij}=-\mathbf{n}_{ji}$\ the unit vector joining the
centers "$i$" and "$j$", after collision $(1-2)$\ velocities of particles
"1" and "2" become%
\begin{equation}
\left\{
\begin{array}{c}
\mathbf{v}_{1}^{\prime }=\mathbf{v}_{1}-\mathbf{n}_{12}\mathbf{n}_{12}\cdot
\mathbf{v}_{1} \\
\mathbf{v}_{2}^{\prime }=\mathbf{n}_{21}\mathbf{n}_{21}\cdot \mathbf{v}_{1}=%
\mathbf{n}_{12}\mathbf{n}_{12}\cdot \mathbf{v}_{1}%
\end{array}%
\right. .  \label{caso-1-1}
\end{equation}%
The second binary collision $(2-3)$\ then brings the velocities of particle
"2" and "3" respectively to%
\begin{equation}
\left\{
\begin{array}{c}
\mathbf{v}_{2}^{\prime \prime }=\mathbf{v}_{2}^{\prime }-\mathbf{n}_{23}%
\mathbf{n}_{23}\cdot \left[ \mathbf{v}_{2}^{\prime }+\mathbf{v}_{1}\right] =%
\mathbf{n}_{12}\mathbf{n}_{12}\cdot \mathbf{v}_{1}-\mathbf{n}_{23}\mathbf{n}%
_{23}\cdot \left[ \mathbf{n}_{12}\mathbf{n}_{12}\cdot \mathbf{v}_{1}+\mathbf{%
v}_{1}\right] \\
\mathbf{v}_{3}^{\prime \prime }=-\mathbf{v}_{1}-\mathbf{n}_{32}\mathbf{n}%
_{32}\cdot \left[ -\mathbf{v}_{1}-\mathbf{v}_{2}^{\prime }\right] =-\mathbf{v%
}_{1}+\mathbf{n}_{23}\mathbf{n}_{23}\cdot \left[ \mathbf{v}_{1}+\mathbf{n}%
_{12}\mathbf{n}_{12}\cdot \mathbf{v}_{1}\right]%
\end{array}%
\right. .  \label{caso-1-2}
\end{equation}%
Finally, the third collision $(1-2)$\ leads particles "1" and "2" to%
\begin{equation}
\left\{
\begin{array}{c}
\mathbf{v}_{1}^{\prime \prime \prime }=\mathbf{v}_{1}^{\prime }-\mathbf{n}%
_{12}\mathbf{n}_{12}\cdot \left[ \mathbf{v}_{1}^{\prime }-\mathbf{v}%
_{2}^{\prime \prime }\right] \\
\mathbf{v}_{2}^{\prime \prime \prime }=\mathbf{v}_{2}^{\prime \prime }-%
\mathbf{n}_{21}\mathbf{n}_{21}\cdot \left[ \mathbf{v}_{2}^{\prime \prime }-%
\mathbf{v}_{1}^{\prime }\right] =\mathbf{v}_{2}^{\prime \prime }-\mathbf{n}%
_{12}\mathbf{n}_{12}\cdot \left[ \mathbf{v}_{2}^{\prime \prime }-\mathbf{v}%
_{1}^{\prime }\right]%
\end{array}%
\right. ,  \label{caso-1-3}
\end{equation}%
namely%
\begin{equation}
\left\{
\begin{array}{c}
\mathbf{v}_{1}^{\prime \prime \prime }=\mathbf{v}_{1}-\mathbf{n}_{12}\mathbf{%
n}_{12}\cdot \mathbf{v}_{1}-\mathbf{n}_{12}\mathbf{n}_{12}\cdot \left\{
\mathbf{v}_{1}-\mathbf{n}_{12}\mathbf{n}_{12}\cdot \mathbf{v}_{1}-\mathbf{n}%
_{12}\mathbf{n}_{12}\cdot \mathbf{v}_{1}+\mathbf{n}_{23}\mathbf{n}_{23}\cdot %
\left[ \mathbf{n}_{12}\mathbf{n}_{12}\cdot \mathbf{v}_{1}+\mathbf{v}_{1}%
\right] \right\} \\
\mathbf{v}_{2}^{\prime \prime \prime }=\mathbf{n}_{12}\mathbf{n}_{12}\cdot
\mathbf{v}_{1}-\mathbf{n}_{23}\mathbf{n}_{23}\cdot \left[ \mathbf{n}_{12}%
\mathbf{n}_{12}\cdot \mathbf{v}_{1}+\mathbf{v}_{1}\right] -\mathbf{n}_{12}%
\mathbf{n}_{12}\cdot \left\{ \mathbf{n}_{12}\mathbf{n}_{12}\cdot \mathbf{v}%
_{1}-\mathbf{n}_{23}\mathbf{n}_{23}\cdot \left[ \mathbf{n}_{12}\mathbf{n}%
_{12}\cdot \mathbf{v}_{1}+\mathbf{v}_{1}\right] -\mathbf{v}_{1}+\mathbf{n}%
_{12}\mathbf{n}_{12}\cdot \mathbf{v}_{1}\right\}%
\end{array}%
\right. .
\end{equation}

\subsubsection{Second case: sequence (2-3)+(1-2)+(2-3)}

Let us now consider the case of the second sequence (\ref{case B}).

After the first binary collision $(2-3)$\ it follows that%
\begin{equation}
\left\{
\begin{array}{c}
\mathbf{v}_{3}^{\prime }=\mathbf{v}_{3}-\mathbf{n}_{32}\mathbf{n}_{32}\cdot
\mathbf{v}_{3}=-\mathbf{v}_{1}+\mathbf{n}_{32}\mathbf{n}_{32}\cdot \mathbf{v}%
_{1} \\
\mathbf{v}_{2}^{\prime }=\mathbf{n}_{23}\mathbf{n}_{23}\cdot \mathbf{v}_{3}=-%
\mathbf{n}_{23}\mathbf{n}_{23}\cdot \mathbf{v}_{1}%
\end{array}%
\right. .  \label{caso-2-1}
\end{equation}%
Then, collision $(1-2)$\ yields%
\begin{equation}
\left\{
\begin{array}{c}
\mathbf{v}_{1}^{\prime \prime }=\mathbf{v}_{1}-\mathbf{n}_{12}\mathbf{n}%
_{12}\cdot \left[ \mathbf{v}_{1}-\mathbf{v}_{2}^{\prime }\right] \\
\mathbf{v}_{2}^{\prime \prime }=\mathbf{v}_{2}^{\prime }-\mathbf{n}_{21}%
\mathbf{n}_{21}\cdot \left[ \mathbf{v}_{2}^{\prime }-\mathbf{v}_{1}\right]%
\end{array}%
\right. ,  \label{caso-2-2}
\end{equation}%
or
\begin{equation}
\left\{
\begin{array}{c}
\mathbf{v}_{1}^{\prime \prime }=\mathbf{v}_{1}-\mathbf{n}_{12}\mathbf{n}%
_{12}\cdot \left[ \mathbf{v}_{1}+\mathbf{n}_{23}\mathbf{n}_{23}\cdot \mathbf{%
v}_{1}\right] \\
\mathbf{v}_{2}^{\prime \prime }=-\mathbf{n}_{23}\mathbf{n}_{23}\cdot \mathbf{%
v}_{1}-\mathbf{n}_{21}\mathbf{n}_{21}\cdot \left[ -\mathbf{n}_{23}\mathbf{n}%
_{23}\cdot \mathbf{v}_{1}-\mathbf{v}_{1}\right]%
\end{array}%
\right. .
\end{equation}%
Finally, collision $(2-3)$\ yields%
\begin{equation}
\left\{
\begin{array}{c}
\mathbf{v}_{2}^{\prime \prime \prime }=\mathbf{v}_{2}^{\prime \prime }-%
\mathbf{n}_{23}\mathbf{n}_{23}\cdot \left[ \mathbf{v}_{2}^{\prime \prime }-%
\mathbf{v}_{3}^{\prime }\right] \\
\mathbf{v}_{3}^{\prime \prime \prime }=\mathbf{v}_{3}^{\prime }-\mathbf{n}%
_{32}\mathbf{n}_{32}\cdot \left[ \mathbf{v}_{3}^{\prime }-\mathbf{v}%
_{2}^{\prime \prime }\right]%
\end{array}%
\right. ,  \label{caso-2-3}
\end{equation}%
namely:%
\begin{equation}
\left\{
\begin{array}{c}
\mathbf{v}_{2}^{\prime \prime \prime }=-\mathbf{n}_{23}\mathbf{n}_{23}\cdot
\mathbf{v}_{1}-\mathbf{n}_{21}\mathbf{n}_{21}\cdot \left[ -\mathbf{n}_{23}%
\mathbf{n}_{23}\cdot \mathbf{v}_{1}-\mathbf{v}_{1}\right] -\mathbf{n}_{23}%
\mathbf{n}_{23}\cdot \left\{ -\mathbf{n}_{23}\mathbf{n}_{23}\cdot \mathbf{v}%
_{1}-\mathbf{n}_{21}\mathbf{n}_{21}\cdot \left[ -\mathbf{n}_{23}\mathbf{n}%
_{23}\cdot \mathbf{v}_{1}-\mathbf{v}_{1}\right] +\mathbf{v}_{1}-\mathbf{n}%
_{32}\mathbf{n}_{32}\cdot \mathbf{v}_{1}\right\} \\
\mathbf{v}_{3}^{\prime \prime \prime }=-\mathbf{v}_{1}+\mathbf{n}_{32}%
\mathbf{n}_{32}\cdot \mathbf{v}_{1}-\mathbf{n}_{32}\mathbf{n}_{32}\cdot
\left\{ -\mathbf{v}_{1}+\mathbf{n}_{32}\mathbf{n}_{32}\cdot \mathbf{v}_{1}+%
\mathbf{n}_{23}\mathbf{n}_{23}\cdot \mathbf{v}_{1}+\mathbf{n}_{21}\mathbf{n}%
_{21}\cdot \left[ -\mathbf{n}_{23}\mathbf{n}_{23}\cdot \mathbf{v}_{1}-%
\mathbf{v}_{1}\right] \right\}%
\end{array}%
\right. .
\end{equation}

Let us compare the outgoing velocities in the two cases. Thus, for example,
in the case of the first sequence:%
\begin{equation}
\mathbf{v}_{1}^{\prime \prime \prime }=\mathbf{v}_{1}-\mathbf{n}_{12}\mathbf{%
n}_{12}\cdot \mathbf{v}_{1}-\mathbf{n}_{12}\mathbf{n}_{12}\cdot \left\{
\mathbf{v}_{1}-\mathbf{n}_{12}\mathbf{n}_{12}\cdot \mathbf{v}_{1}-\mathbf{n}%
_{12}\mathbf{n}_{12}\cdot \mathbf{v}_{1}+\mathbf{n}_{23}\mathbf{n}_{23}\cdot %
\left[ \mathbf{n}_{12}\mathbf{n}_{12}\cdot \mathbf{v}_{1}+\mathbf{v}_{1}%
\right] \right\} .  \label{first case-of-v1-}
\end{equation}%
As a result:%
\begin{equation}
\mathbf{v}_{1}^{\prime \prime \prime }=\mathbf{v}_{1}-\mathbf{n}_{23}\mathbf{%
n}_{23}\cdot \left[ \mathbf{n}_{12}\mathbf{n}_{12}\cdot \mathbf{v}_{1}+%
\mathbf{v}_{1}\right] ,  \label{test-first case-of-v1}
\end{equation}%
while in the second sequence one obtains%
\begin{equation}
\mathbf{v}_{1}^{\prime \prime }=\mathbf{v}_{1}-\mathbf{n}_{12}\mathbf{n}%
_{12}\cdot \left[ \mathbf{v}_{1}+\mathbf{n}_{23}\mathbf{n}_{23}\cdot \mathbf{%
v}_{1}\right] .  \label{test-second case-of-v1}
\end{equation}

It therefore follows that the final velocities $\left\{ \mathbf{v}%
_{1}^{(f)}(t_{i}),\mathbf{v}_{2}^{(f)}(t_{i}),\mathbf{v}_{3}^{(f)}(t_{i})%
\right\} $ are generally different in the two cases (\ref{case A}) and (\ref%
{case B}), except if\textbf{\ }$\mathbf{n}_{12}(t_{i})\cdot \mathbf{n}%
_{23}(t_{i})=0,1$. This implies that double collision laws cannot be
uniquely defined in terms of sequences of binary elastic collisions, i.e.,
in other words that double collision laws are ill defined.

\subsection{3.5 - Indirect proof of Propositions \#3}

An indirect proof of\ Propositions \#3 can also be obtained from the proof
of Proposition \#2 given above. For this, it is sufficient to notice that,
in validity of the condition of symmetry (\ref{symmetry}), then one obtains
respectively%
\begin{equation}
\mathbf{v}_{1}^{\prime \prime \prime }=\mathbf{v}_{1}-2\mathbf{n}_{12}%
\mathbf{n}_{12}\cdot \mathbf{v}_{1},
\end{equation}%
and similarly%
\begin{equation}
\mathbf{v}_{3}^{\prime \prime \prime }=-\mathbf{v}_{1}+2\mathbf{n}_{12}%
\mathbf{n}_{12}\cdot \mathbf{v}_{1}\mathbf{.}
\end{equation}

\section{4 - Basic implications: ill-defined and stochastic B-S systems}

We intend to show that the previous propositions imply fundamental
consequences for the physical interpretation of the B-S system. More
precisely, depending on how multiple collisions laws are prescribed, two
possible\ cases can be distinguished. The first one is the case in which
multiple collisions remain actually what they appear, namely non-unique and
therefore ill-defined.

This is a consequence of Propositions \#1 and \#2, which actually imply\
three main conclusions. The first two are as follows:

\subsection{4.1 - Conclusion \#1:\ the ordinary B-S system is ill-defined}

There is no unique deterministic law which can be associated with general
multiple collisions occurring in the B-S system (i.e., which correspond to
arbitrary admissible initial conditions, namely compatible with the possible
physical constraints set on particle motion). Therefore, in the presence of
multiple collisions, the B-S system remains generally (i.e., except special
"symmetric" initial conditions) ill-defined, which means that the time
evolution of the same B-S system is not uniquely defined. In such a case the
B-S system will be referred to as\textbf{\emph{\ }}\emph{ordinary}\textbf{%
\emph{\ }}(O-B-S system). However, there is a physically-grounded
alternative, which occurs if a suitable set of stochastic collision laws is
introduced for multiple collisions.

\subsection{4.2 - Conclusion \#2: stochastic character of multiple collisions%
}

As shown above, multiple collisions correspond, however, to non-unique, but
well-defined, "events", namely ordered sequences of simultaneous binary
collisions (see in particular Eqs.(\ref{test-first case-of-v1}) and (\ref%
{test-second case-of-v1}) in the case of double collisions). This means that
multiple collisions, and the same ordered sequences by which they are
realized, can be considered\ stochastic, namely they are not only 1)
non-unique but also 2) endowed with a well-defined probability density,
namely they are equiprobable.\

The reason why condition 2) occurs is that in a multiple (i.e., in the case
considered here, binary) collision there is no way to decide which of the
sequences actually happens. Moreover, since multiple collisions occur among
like particles, it is obvious that all ordered sequences of simultaneous
binary collisions must have the same probability density to happen, i.e.,
because they cannot depend on the permutation of the labeling indexes of the
colliding particles. Hence, the probability densities are permutation
symmetric. Based on these considerations, the following proposition holds.

\textbf{PROPOSITION \#4: stochastic character of generally non-unique
multiple collisions}

\emph{When multiple collision laws are non-unique, then they are necessarily
stochastic in character, i.e.:}

\emph{1) They can be represented by different sequences of instantaneous
binary collisions.}

\emph{2) Each sequence of instantaneous binary collisions, which occurs in a
multiple collision,\textbf{\ }can be assigned a\textbf{\ }suitable
probability.}

\emph{3)} \emph{All sequences\ of instantaneous binary collisions occurring
in the same multiple collision are equally probable.}

\bigskip

\emph{PROOF OF PROP. \#4}

The proof is straightforward. In fact, as shown above (PROP. \#2 and \#3),
multiple collisions are generally characterized by non-unique collisions
laws which correspond to distinct ordered sequences of instantaneous binary
collisions. Therefore this implies that multiple collision laws are
generally non-unique, and because of permutation symmetry, also stochastic.
This yields also the following third conclusion.

\subsection{4.3 - Conclusion \#3:\ stochastic B-S system}

If multiple collision laws are considered stochastic, i.e., defined by means
of stochastic multiple-collision laws (namely in which each possible\
ordered sequence of binary collisions is considered stochastic and equally
probable)\ then the time evolution of the B-S system is stochastic.

In the remainder we shall therefore distinguish the two cases in which
multiple collisions are considered either ill defined or stochastic, namely
respectively the ordinary and the stochastic B-S systems.

Based on the previous arguments the main result is then provided by the
following corollary.

\bigskip

\textbf{COROLLARY TO PROP. \#4: multiple-collision laws of the stochastic
B-S system}

\emph{For the stochastic Boltzmann-Sinai system (S-B-S system), the
collision laws for multiple collisions \emph{are generally stochastic and
are provided by all possible sequences of\ instantaneous binary collisions,
with the different se}quences to be regarded as equally probable.}

\section{5 - The non-classical $2-$way logic of the ordinary B-S system and
the $3-$way logic of the stochastic B-S system}

We are now ready to establish the propositional logic appropriate for the
two systems considered here, namely respectively the O-B-S and S-B-S systems.

\subsection{5.1 - The logic of the O-B-S system}

It is immediate to realize that for the O-B-S system the logic remains the
same one which applies to Newton's Classical Cosmology (see subsection 2.1).
In fact, the only possible logical truth propositions which therefore define
the semantics of the O-B-S system are:

\begin{itemize}
\item \textbf{1st logical truth proposition "\emph{true"}: }the truth
logical value\textbf{\ "}\emph{true}\textbf{"} for the B-S system is
represented by the "phase flow" generated by the same system, whose
existence is warranted "a priori" in the time interval $I$ ($\subseteq
\mathbb{R}
)$ provided no multiple collisions occurs in the same time interval. Hence,
the unique phase flow, i.e., the phase mapping, defined for all $t,t_{o}\in
I $ ($\subseteq
\mathbb{R}
)$,
\begin{equation}
\mathbf{x}(t_{o})\equiv \mathbf{x}_{o}\rightarrow \mathbf{x}(t)\equiv \chi (%
\mathbf{x}_{o},t_{o},t),
\end{equation}%
relates a prescribed non-multiple collision state $\mathbf{x}(t_{o})\equiv
\mathbf{x}_{o}\equiv (\mathbf{r}(t_{o})=\mathbf{r}_{o},\mathbf{v}(t_{o})=%
\mathbf{v}_{o})$ with its non-collisional image at time $t$ determined by
the same phase flow, namely $\mathbf{x}(t)\equiv (\mathbf{r}(t),\mathbf{v}%
(t))\equiv \chi (\mathbf{x}_{o},t_{o},t)$ which is prescribed at any other
time $t\neq t_{o}$.

\item \textbf{2nd logical truth proposition "\emph{false"}:} the truth
logical value\textbf{\ "}\emph{false}\textbf{" }for the same system is
represented, instead, by any other mapping of the type\textbf{\ }%
\begin{equation}
\mathbf{x}(t_{o})\equiv \mathbf{x}_{o}^{\prime }\rightarrow \mathbf{x}%
^{\prime }(t),
\end{equation}%
where for all times\ $t\neq t_{o},$\ $x^{\prime }(t)\equiv \chi
(x_{o}^{\prime },t_{o},t)$\ and\ $x_{o}^{\prime }\neq x_{o}$\ is an
arbitrary initial state differing from $x_{o}$. Notice that $x_{o}^{\prime }$%
\ may include also non-accessible initial states for which the phase flow is
not defined.

\item As a consequence, the ordinary B-S system formally satisfies the
customary form of the $2-$way Aristotle's \emph{Principle of
non-contradiction}, namely:
\end{itemize}

\bigskip

\textbf{Theorem 1 - The two fundamental logical propositions of the
O-B-S-system (standard) logic (Aristotle's 2-WAY PNC)}

\emph{Given the N-body ordinary B-S system (O-B-S system) defined above, the
following fundamental logical and mutually exclusive logical propositions
apply:}

\emph{\textbullet\ 1st proposition: classical truth.}

\emph{\textbullet\ 2nd proposition: classical falsehood.}

\bigskip

\subsection{5.2 - The 3-way logic of the S-B-S system}

In case of the S-B-S system instead the truth propositions which apply to
Newton's Classical Cosmology (see subsection 2.1) still apply provided in
the time interval $I$ ($\subseteq
\mathbb{R}
)$ no multiple collision occurs. However, a third truth proposition now
arises explicitly, precisely:

\begin{itemize}
\item \textbf{3rd\ logical truth proposition "\emph{indeterminate"}} (or "%
\textbf{\emph{undecidable"}})\textbf{: }the truth logical value\ \textbf{%
\emph{"\textbf{indeterminate}", }}which may be equivalently called \emph{"%
\textbf{undecidable}"}, for the same system is represented, instead, by any
other mapping of the type%
\begin{equation}
\mathbf{x}(t_{o})\equiv \mathbf{x}_{o}\rightarrow \mathbf{x}^{\prime
}(t)\equiv \chi _{S}(\mathbf{x}_{o},t_{1},t),
\end{equation}%
where $\chi _{S}(\mathbf{x}_{o},t_{1},t)$ can be intended either as a
suitable stochastic or multiple-valued function defined for any time $%
t_{1}\geq t_{o},$\ $t\neq t_{o},$\ $x^{\prime }(t)\equiv (r^{\prime
}(t),v^{\prime }(t))$, which takes into account the presence of multiple
collisions occurring during the same time interval and represents all
possible accessible states occurring for the same S-B-S system.

\item As a consequence, the semantics of the S-B-S system, namely its
logical truth values now satisfies the $3-$\emph{way Principle of
non-contradiction, }namely:
\end{itemize}

\bigskip

\textbf{Theorem 2 ---The three fundamental logical propositions of the
non-standard CM logic (3-WAY PNC)}

\emph{Given the N-body stochastic B-S system (S-B-S system) defined above,
the following fundamental logical and mutually exclusive logical
propositions apply:}

\emph{\textbullet\ 1st proposition: classical truth.}

\emph{\textbullet\ 2nd proposition: classical falsehood.}

\emph{\textbullet\ 3rd proposition: classical undecidability.}

\emph{Proof:}\textbf{\ }In fact, unlike in CL, for the S-B-S system the
deterministic principle does not hold any more (this happens after the
occurrence of any multiple collision). As a consequence three possible truth
values exist, namely truth, falsehoods and undecidability (which may be
termed also as classical because they refer to CM). Thus, for example, truth
is (respectively falsehood and undecidability) is the common property among
all possible true (false, undeterminate) phase-space trajectories. Finally,
by construction the three logical values are manifestly mutually
incompatible. As a consequence a 3-way Principle of non-contradiction (3-way
PNC) holds.\textbf{\ Q.E.D.}

\bigskip

Therefore, the theorems indicated above (Thms. 1 and 2) identify
respectively the standard 2-way classical logic and the (non-standard) 3-way
non-classical logic which are based on the occurrence of two and,
respectively, three different truth logical values.

The result obtained by Theorem 2 proves, in particular, that the logic of CM
must be non-standard and based on the 3-way PNC.

\subsection{5.3 - Physical implications}

It has to be noted that, in difference with respect to the S-B-S system
which is defined globally in time (i.e., for all times $t\in I\equiv
\mathbb{R}
$),\ the time evolution of the O-B-S system can only be prescribed for all
times smaller than the minimum critical time $t_{\ast }$.\ For this reason,
in principle, the question of the physical relevance of the O-B-S system
might be called into question. On the other hand, there is no way to predict
or to exclude possible multiple collision events. For this reason it follows
that, at least in principle, the only physical identification for the B-S
system which makes sense from the physical standpoint (in the sense that it
is the only one which is globally defined), is only that of identifying it
with the S-B-S system.

This means also that, in the present context, the notion of the \emph{%
undecidability}\textbf{\ }(or \emph{indeterminacy})\textbf{\ }truth logical
value is applicable only to the same S-B-S system.

Regarding the consistency of such a notion with that available in
mathematics, i.e., specifically in the context of G\"{o}del
completeness/incompleteness theorems \cite{Goedel}), we notice that the
undecidability\ logical truth value in the context the S-B-S system arises
if the application phase mapping
\begin{equation}
t\rightarrow \mathbf{x}(t)=\chi (\mathbf{x}_{o},t_{o},t)
\end{equation}%
(with $x(t)$ denoting the state at time $t,$ for the initial condition $%
x(t_{o})=x_{o}$ and $\chi (x_{o},t_{o},t)$ denoting the phase flow) is
stochastic at time $t>t_{o}$, i.e., provided a (non-symmetric) multiple
collision (at, say, a time $t_{1}$) occurs in the open time interval $\left(
t_{o},t\right) $. This means that for any $t>$\ $t_{1}$\ the flow $\chi
(x_{o},t_{o},t)$\ become stochastic and therefore that the state $x(t)$ is
necessarily discontinuous at time $t_{1}$ in the velocity\ $v(t)$. \ This
also implies that the notion of undecidability makes sense also from the
mathematical viewpoint since it corresponds to a well-definite mathematical
expression, i.e., the value of velocity jump across the discontinuity, whose
value cannot be uniquely determined. The implication is therefore that the
S-B-S system represents an example-case in which incompleteness occurs, and
therefore an example-case in which G\"{o}del theory applies. \

The notable aspect is that this happens - namely the logical truth value
undecidability occurs - despite the fact that\textbf{\ }apparently\textbf{\ }%
the configuration space $\Omega $ for the hard-sphere system seems to
coincide with a bounded subset of the $3$-dimensional Euclidean space $%
\mathbb{R}
^{3}$\ (a case in which, according to G\"{o}del, undecidability should be
ruled out). The explanation, however, is immediate and follows by proving
that the actual configuration space has necessarily a higher dimension in
the presence of multiple collisions. In fact the configuration space is, by
definition, the set in which position trajectories are uniquely mapped.
Therefore in the S-B-S system, if $j$\ multiple trajectories exist in $%
\Omega $. the actual configuration space coincides necessarily with the
extended configuration space $\Omega ^{f}$ represented by the product $%
\Omega ^{f}=\prod\limits^{i=1,f}\Omega $, i.e. a subset of the product space
$\left( E^{3}\right) ^{f}.$

Therefore, this conclusion proves the connection with G\"{o}del theory \cite%
{Goedel}: in other words the undecidable propositions arising in the S-B-S
system are consistent with the same theorems. The actual physical mechanism
which gives rise to the occurrence of such logical propositions is produced
by multiple collisions. This confirms therefore that the B-S system
represents an example-case and an application of G\"{o}del undecidability
theorems.

The next physically-relevant issue which one might wish to address concerns
therefore the question of "how likely" are multiple collision events in the
S-B-S (or O-B-S) system \cite{saari71,saari73,saari98,Pollard-Saari,Knauf}.
This amounts to estimate the relative (canonical or microcanonical)
invariant measure of the subset of phase-space in which multiple collisions
events occur. As shown in the following subsection, such a measure can be
estimated and shown to be identically vanishing.\ Its invariance property
warrants that the same measure remains always vanishing at all times. In
other words, excluding a subset of phase space with vanishing measure, all
possible initial conditions for the hard spheres (which belong to the S--B-S
system) lead\ in an arbitrary finite time interval to ordinary binary
collisions events only.

\subsection{5.4 - Estimate of the relative microcanonical measure of the
subset of multiple collisions}

As is well-known the microcanonical measure of the B-S system is represented
by an integral on the energy hypersurface. Thus, for a system formed by $N>2$
smooth hard spheres of diameter $\sigma $, particle kinetic energies $%
E^{(i)}=\frac{1}{2}mv^{(i)2}$ (for $i=1,N$ ) and with total kinetic energy
of the hard spheres%
\begin{equation}
E_{N}=\sum\limits_{i=1,N}E^{(i)},
\end{equation}%
such a measure is represented by the integral%
\begin{equation}
\Omega (\alpha )=\int\limits_{\Gamma _{N}}dx\overline{\Theta }^{(N)}(\mathbf{%
r)}\delta \mathbf{(}E_{N}-\alpha ),
\end{equation}%
which is performed on the "admissible subset" (the subset of admissible
configurations of the $N-$body phase-space $\Gamma _{N}$),\ namely the
subset of $\Gamma _{N}$\ where $\overline{\Theta }^{(N)}(r)=1$, which
coincides with the subset of a parallepiped-shaped bounded subset of the
Euclidean space $\mathbb{R} ^{3}$, i.e., the billiard, which can be occupied
by the hard spheres. Here the notation is standard \cite{CMFD-T-1}. Thus, $dx
$ is the volume element $dx=\prod\limits_{i,j=1,N}d^{3}r_{i}d^{3}v_{j}$, $%
\Theta _{N}(\mathbf{r})$ denotes the "ensemble theta function", which by
definition is set equal to unity on the said admissible subset of
configuration space. Thus, $\overline{\Theta }^{(N)}(\mathbf{r)}$ is defined
as \cite{CMFD-T-1}:%
\begin{equation}
\overline{\Theta }^{(N)}(\mathbf{r)}=\sum\limits_{\substack{ i,j=1,N  \\ %
(i<j)}}\overline{\Theta }(\left\vert \mathbf{r}_{i}-\mathbf{r}%
_{j}\right\vert -\sigma \mathbf{)}\sum\limits_{k=1,N}\overline{\Theta }%
(\left\vert \mathbf{r}_{k}-\frac{\sigma }{2}\mathbf{n}_{k}\right\vert -\frac{%
\sigma }{2}\mathbf{),}
\end{equation}%
where $\overline{\Theta }(x\mathbf{)}$ denotes the strong Heaviside theta
function%
\begin{equation}
\overline{\Theta }(x\mathbf{)=}\left\{
\begin{array}{c}
1\text{ if }x\geq 0 \\
0\text{ if }x<0%
\end{array}%
\right. \text{,}
\end{equation}%
and $\mathbf{n}_{k}$ is the inward unit vector orthogonal to the boundary at
the point of contact of the $k-$th hard sphere having baricentric position $%
\mathbf{r}_{k}$.

Next let us pose the problem of estimating the possible effect of multiple
collisions, such as for example double collisions. The proof is elementary
and is based on the definition of the ensemble theta function $\overline{%
\Theta }^{(N)}(\mathbf{r)}$ given above.

Clearly, this must correspond to a subset of the collisional subset of
binary collisions because double collisions occur when two binary collisions
in turn occur, which involve the same particles. For convenience let us
therefore introduce a measure on the binary collision subset. This can be
defined in terms of a phase-space integral in which binary collisions are
specifically taken into account, namely are prescribed in terms of Dirac
deltas of the type $\delta (\left\vert \mathbf{r}_{i}-\mathbf{r}%
_{j}\right\vert -\sigma \mathbf{)}$ (for $i,j=1,N$), i.e.,%
\begin{equation}
\Omega _{2}(\alpha )=\int\limits_{\Gamma _{N}}dx\overline{\Theta }^{(N)}(%
\mathbf{r)}\delta \mathbf{(}E_{N}-\alpha)\sum\limits_{i,j=1,N;i<j}\delta
(\left\vert \mathbf{r}_{i}-\mathbf{r}_{j}\right\vert -\sigma \mathbf{)}.
\end{equation}%
Then, thanks to the assumption of like particles, this implies that $\Omega
_{2}(\alpha )$ must necessarily take the form%
\begin{equation}
\Omega _{2}(\alpha )=N(N-1)\int\limits_{\Gamma _{N}}dx\overline{\Theta }%
^{(N)}(\mathbf{r)}\delta \mathbf{(}E_{N}-\alpha )\delta (\left\vert \mathbf{r%
}_{1}-\mathbf{r}_{2}\right\vert -\sigma \mathbf{),}
\end{equation}%
where the Dirac delta takes into account the contribution of binary
collisions (1-2), i.e., occurring between particle "1" and "2", only. Hence
the multiple integral evaluated on the $2-$body phase space $\Gamma _{2}$
can always be evaluated explicitly to yield
\begin{equation}
\int\limits_{\Gamma _{2}}d\mathbf{x}_{1}d\mathbf{x}_{2}\overline{\Theta }%
^{(N)}(\mathbf{r)}\delta \mathbf{(}E_{N}-\alpha )\delta (\left\vert \mathbf{r%
}_{1}-\mathbf{r}_{2}\right\vert -\sigma \mathbf{)}=\int\limits_{\Gamma _{2}}d%
\mathbf{v}_{1}d\mathbf{r}_{1}d\Sigma _{21}d\mathbf{v}_{2}\overline{\Theta }%
^{(N)}(\mathbf{r)}\delta \mathbf{(}E_{N}-\alpha ),  \label{last}
\end{equation}%
where $\Sigma _{21}$ represents the integration on a suitable solid angle
element.

Now, in principle, the integral on the rhs of Eq.(\ref{last}) might still
depend also on multiple collisions (to be taken into account through the
factor $\overline{\Theta }^{(N)}(\mathbf{r)}$), such as for example, double
collisions in which either particle "1" or "2" collide with a third particle
"3", namely corresponding to possible binary collisions of the type (1-3)
and (2-3). We intend to demonstrate that such contributions (to $\Omega
_{2}(\alpha )$) must vanish identically. The elementary proof is as follows.

First, one notices that in the integral on the rhs, the ensemble theta
function $\overline{\Theta }^{(N)}(\mathbf{r)}$ can always be replaced with
a smooth function $\overline{\Theta }^{(N)}(\mathbf{r,}\varepsilon ^{2}%
\mathbf{)}$ such that%
\begin{equation}
\lim_{\varepsilon ^{2}\rightarrow 0}\overline{\Theta }^{(N)}(\mathbf{r,}%
\varepsilon ^{2}\mathbf{)}=\overline{\Theta }^{(N)}(\mathbf{r)},
\end{equation}%
in which $\overline{\Theta }^{(N)}(\mathbf{r,}\varepsilon ^{2})$ can be
taken of the type
\begin{equation}
\overline{\Theta }^{(N)}(\mathbf{r,}\varepsilon ^{2})\propto \Theta
(\left\vert \mathbf{r}_{1}-\mathbf{r}_{3}\right\vert -\sigma +\varepsilon
^{2}\mathbf{)}\Theta (\sigma +\varepsilon ^{2}-\left\vert \mathbf{r}_{1}-%
\mathbf{r}_{3}\right\vert )\Theta (\left\vert \mathbf{r}_{2}-\mathbf{r}%
_{3}\right\vert -\sigma +\varepsilon ^{2}\mathbf{)}\Theta (\sigma
+\varepsilon ^{2}-\left\vert \mathbf{r}_{2}-\mathbf{r}_{3}\right\vert ).
\end{equation}%
Now it is straightforward to show, by direct evaluation of the same
phase-space integrals, that multiple collisions of the type (1-2)+(1-3) or
(1-2)+(2-3) produce contributions at most of order $\mathbf{\varepsilon }%
^{2} $ to $\Omega _{2}(\alpha )$. As a consequence, once the limit $%
\varepsilon ^{2}\rightarrow 0$ is taken, this means that double collisions
can only give rise to vanishing contributions to $\Omega _{2}(\alpha )$.
Analogous considerations can be made for higher-order multiple collisions.
This proves therefore that multiple collisions give vanishing contributions
to the microcanonical measure of the collisional subset.

\section{6 - Conclusions: the logic of classical mechanics}

The very concept of logic attached to classical mechanics (CM) is
traditionally rooted in the notion of classical logic. Goal of this paper
has been to challenge the deterministic logic of classical mechanics,
showing that a 3-way non-standard logic actually holds which should be
therefore considered valid for the whole set of classical physics. However,
another striking conclusion arises. In fact, based on previous
investigations on the logic of quantum physics (quantum mechanics and
quantum gravity), the same unique logical principle of non-contradiction and
the same discrete truth values appear to encompass both classical and
quantum physics.

To achieve such a result, classical mechanics should include classical
systems which are intrinsically stochastic.

For this purpose, in this paper the logic of classical mechanics has been
tackled for the Boltzmann-Sinai (B-S) system formed by $N>2$\ smooth hard
spheres subject to multiple elastic collisions. Multiple collisions have
been shown to be generally realized by different sequences of simultaneous
binary collision events, which implies that on the same grounds multiple
collision laws are ill defined. Here, as an illustration, the case of double
collisions has been considered.

To overcome the difficulty, however, suitable stochastic collision laws have
been adopted for the B-S hard-sphere system, which has been referred to for
this reason as stochastic B-S system.

Its logic - and therefore also that of CM - has been shown to be a
non-standard logic, in the sense that it includes a third logical truth
value,\ referred to as indeterminate/undecidable.\ In particular such a
logical value has been shown to be compatible with the notion of
undecidability adopted in the context of G\"{o}del theory \cite{Goedel}. As
a consequence, this means that the stochastic Boltzmann-Sinai system (S-B-S
system) represents an example-case of undecidability according to G\"{o}del
undecidability theorems \cite{Goedel}.

Thus, it coincides necessarily with that of a $3-$valued propositional
calculus. In other words:

\begin{itemize}
\item The logic of CM has been shown to be non classical and to coincide
with the non-standard logic which is governed by the so-called $3-$\emph{way
Principle of non-contradiction}\textbf{\ }\emph{(3-way PNC).}

\item Such a logical principle coincides with the $3-$way logical principle
previously shown to hold for quantum physics, i.e., both for
non-relativistic quantum mechanics (QM) and quantum gravity (QG)\ \cite%
{LOGIC-1,LOGIC-2}.

\item Therefore, a unique $3-$way PNC has been shown to govern both
classical and quantum physics.
\end{itemize}

Regarding the comparison between CM and respectively both QM and QG, we
stress that two common features arise, namely 1) the existence in all such
cases of the three logical values "true", "false" and "\textbf{\emph{%
undecidable"}} (or \textbf{"\emph{indeterminate"}});\ 2) the validity in all
such cases of the same 3-way principle of non-contradiction. However, the
precise definitions of the truth values which occur in QM and QG generally
differ from those obtained here for CM, being related to the properties of
appropriate quantum observables, rather than the position 3-vector $\mathbf{r%
}(t)$\ holding in the case of CM.

Nevertheless, the definitions of the same truth values are not independent,
in the sense that they must actually be prescribed in a consistent manner.
In fact, when QM is evaluated in the semiclassical limit then it is possible
to prove that it reduces exactly to CM (see details of the proof given in
Ref. \cite{LOGIC-2}). This implies that, depending on the specific
prescription for the quantum system (which should be identified with the
quantum B-S system), also the same logic of CM obtained here is expected to
be recovered in the semiclassical limit.

These conclusions represent a potential notable innovation in the logical
dichotomy true/false. This is a crucial topic which has crossed millennia
through philosophy, logic, mathematics, physics and even theology. In the
history of philosophy and science, great minds tackled the challenge,
ranging from classic Authors like Confucius (Kong Fuzi), Lao Tzu (Laozi),
Aristotle (Aristot\'{e}l\={e}s), Middle Ages Authors Thomas Aquinas and
Ramon Llull, Age of Enlightenment Authors like Immanuel Kant and modern
Authors like Kurt Friedrich G\"{o}del, George David Birkhoff, John von
Neumann and Alan Turing. In the present research we have connected the
philosophical arguments to the physics of Galileo Galilei, Isaac Newton,
Ludwig Boltzmann, Harold Grad and Yakov Grigorevich Sinai, with special
reference to particle trajectories in the case of hard-spheres undergoing
multiple collisions which are unavoidably stochastic.

\bigskip

\bigskip

\textbf{Data Availability Statement}

This manuscript has no associated data. [Authors' comment: The paper reports
theoretical analytical study. All data pertinent to this study are contained
in the paper.]

\textbf{Code Availability Statement}

This manuscript has no associated code/software. [Authors' comment: The
paper reports theoretical analytical study and no use of code/software is
made.]

\textbf{Declaration of competing interest}

The authors declare that to their knowledge there are no competing financial
interests or personal relationships that could have appeared to influence
the work reported in this paper.

\textbf{Funding information}

This research received no external funding.

\textbf{Conflict of Interest}

The authors declare no conflicts of interests.

\textbf{Author Contributions}

Conceptualization, Massimo Tessarotto, Claudio Cremaschini, Claudio Asci,
Alessandro Soranzo, Marco Tessarotto and Gino Tironi; Investigation, Massimo
Tessarotto, Claudio Cremaschini, Claudio Asci, Alessandro Soranzo, Marco
Tessarotto and Gino Tironi; Writing -- original draft, Massimo Tessarotto,
Claudio Cremaschini, Claudio Asci, Alessandro Soranzo, Marco Tessarotto and
Gino Tironi. All authors have read and agreed to the published version of
the manuscript.

\end{document}